# Wave-Particle Behaviour in Bridge Theory

*Massimo AUCI*
*Space Science Department*
*Odisseo Space*
*Via Battistotti Sassi 13, 20133 Milano, Italy*

The Bridge Theory that is based on the role that the transversal component of the Pointing vector has in the localisation in the neighbourhood of a dipole of an amount of energy and momentum equal to ones of a photon of same frequency, if applied at the collision of a pair of real particles is able to describe their travel in space in agreement with the Relativistic Quantum Theory. This result suggests that Relativity and Quantum Mechanics despite their apparent incompatibility they become compatible if we assume the existence of a deepest level of reality in which is the electromagnetism to produce the quantum and relativistic effects in the matter allowing in the same context a dualism wave-particle equivalent to that hypothesized by Louis de Broglie.

**1) Introduction**

The Bridge Theory (BT) (see ref [1]) is a theoretical evolution of the Dipolar Electromagnetic Source model (DEMS) (see ref. [2-3-4]). The theory proves that when two opposite charged particles interact within a same pair with a time of collision T, the corresponding dipolar electromagnetic source localises energy and momentum equal to ones of a photon of frequency $T^{-1}$. The wavelength of the DEMS measured in the CM of the source is equal to the length of the dipole moment, which is equal to the minimum distance achieved during the reciprocal interaction of the two charges.

In ref. [5] we have recently proved that all the observers of a same dipole, if placed in different points of space they observe along their ideal lines of view different relative velocities of approach and a transversal different projections of the length of the dipole moment. In fact, for the energy and momentum conservation laws, the wave observed cannot be seen from all observers with the same energy of the photon $E\gamma$, because the energies, momenta and frequencies observed are described by the three equations:

$$\begin{cases} E = \dfrac{\varepsilon}{\sqrt{1-\beta^2}} \\ E_\gamma = h\nu \\ E_\gamma = \dfrac{E^2 - P^2c^2}{2(E - Pc\cos\theta)} \end{cases}, \qquad (1a)$$

the first describe the energy involved in the formation of the DEMS, the second is the energy of the localised photon, the last is the energy observed as a function of the angular position of the observer. The third equation can be also rewritten in terms of relativistic Doppler frequency for a moving source or symmetrically for moving observers:

$$\nu = \nu_0 \frac{\sqrt{1-\beta^2}}{1-\beta\cos\theta} \qquad (1b)$$

where $\nu_0 = \frac{\varepsilon}{2h}$ is the wave rest frequency. The equation (1b) describes the wave frequency that connects the source with all other observer, this produces an electromagnetic net (see ref. [5]) able to adjust and redistribute energy and momentum between source and observers in space-time in agreement both with the quantum and relativistic phenomenologies

Furthermore we will prove that equations (1b) could be used to give a wave description of matter that corresponds to the de Broglie one. This description is consistent with BT and it is simultaneously in agreement with Quantum Mechanics and Relativity as a complete theory would require.

## 2) Energy and momentum of the DEMS

In order to the symmetry between charge and anti - charge, two observers placed on each of two particles in the respective frames $S_1$ and $S_2$, they can measure reciprocally the total energy of the DEMS produced during their interaction as the total energy carried by each of one of the two particles that impinge one on the other without that the two particles they have informations about the own energy or mass.

To perform simultaneous measures of the energy and momentum of each of two particles, we must consider a third special external frame S that we define the lab frame. To do the measures we need to produce two independents DEMSs. We suppose that S has the same nature of a neutral polarizable field or of the ordinary atomic matter, therefore when a pair crosses the frame S the two particles interact polarising the media with the effect to produce the two non punctual DEMSs[1]:

$$\bullet SS_1 \, , \, \bullet SS_2 \, .$$

From the experimental point of view and in order to represent in terms of mass the energy and momentum of each source, we assume that each real particle has got an own rest energy $\varepsilon = m_0 c^2$ in such a way that, in agreement with BT (see ref. [5]) and with Relativity, for each DEMS we can write:

   i) energy and momentum of the particle <1> observed from the frame S

$$\bullet SS_1 \begin{cases} E_1 = \gamma_1 \, m_{0,1} \, c^2 \\ \mathbf{p}_1 = \gamma_1 \, \boldsymbol{\beta}_1 \, m_{0,1} \, c \end{cases} \qquad (2a)$$

---

[1] We consider each DEMS as a pair of frames and we denote the source with the symbol "●(frame1)(frame2)".

ii) energy and momentum of the particle <2> observed from the frame S

$$\bullet SS_2 \quad \begin{cases} E_2 = \gamma_2\, m_{0,2}\, c^2 \\ \mathbf{p}_2 = \gamma_2\, \boldsymbol{\beta}_2\, m_{0,2} c \end{cases} \qquad (2b)$$

Now in order to calculate the total energy and momentum in the CM of the two interacting particles respect S, we need of the total rest energy of the DEMS $\bullet S_1 S_2$, and of the relativistic factors $\beta$ and $\gamma$.

So we obtain:
- the rest energy of the source

$$\varepsilon_{CM} = \sqrt{(E_1 + E_2)^2 - |\mathbf{p}_1 + \mathbf{p}_2|^2 c^2} \quad ; \qquad (3)$$

- the CM velocity respect the lab frame

$$\boldsymbol{\beta} = \frac{(\mathbf{p}_1 + \mathbf{p}_2)c}{E_1 + E_2} \quad ; \qquad (4)$$

- the gamma factor

$$\gamma = \frac{E_1 + E_2}{\varepsilon_{CM}} \quad ; \qquad (5)$$

that allow to write for the travelling source

$$\bullet S_1 S_2 \quad \begin{cases} E = E_1 + E_2 = \gamma\, \varepsilon_{CM} \\ \mathbf{P} = \mathbf{p}_1 + \mathbf{p}_2 = \gamma \boldsymbol{\beta}\, \dfrac{\varepsilon_{CM}}{c} \end{cases} \qquad (6)$$

**3) Doppler effect and de Broglie wave description for the two DEMSs $\bullet S_1 S_2$ and $\bullet SS_i$**

Since the frame associated to the CM of the source $\bullet S_1 S_2$ is in motion with respect to the lab frame, using the third of the eq. (1a) and the eq. (6) we write:

$$E_\gamma = \tfrac{1}{2} \varepsilon_{CM} \frac{\sqrt{1-\beta^2}}{1 - \beta \cos\theta} \qquad (7)$$

we note that each external observer in S can observe the source as emitting photons with a Doppler energy $E_\gamma$ as a function of the angle $\theta$ between the total momentum $\mathbf{P}$ of the DEMS and the momentum $\mathbf{P}_\gamma$ of the photon observed in S.

Using the Taylor series we also to evidence that for very high energy of collision, i.e. $\beta \cong 1$, the energy emission along each other angle different from 0° can be neglected; whereas for very low energy, i.e. $\beta \cong 0$, the photon can be emitted casually in any direction, in this case its wavelength converges to the Compton one related to the CM of the interacting particles:

$$\lambda_0 = \frac{hc}{\frac{1}{2}\varepsilon_{CM}} \ ,$$

instead for high energy we have a forward very narrow emission of photons along the axis of dipole of the DEMS with energy and momentum:

$$\begin{cases} E_\gamma \approx \frac{1}{2}\gamma\varepsilon_{CM} \dfrac{1+\beta}{1+\dfrac{\beta\theta^2}{2(1-\beta)}} \approx \gamma\varepsilon_{CM} \dfrac{(1+\beta)}{2} \\ P_\gamma \approx \frac{1}{2}\gamma\beta\varepsilon_{CM} \dfrac{1+\dfrac{1}{\beta}}{1+\dfrac{\beta\theta^2}{2(1-\beta)}} \approx \frac{1}{2}\gamma\beta\dfrac{\varepsilon_{CM}}{c}\dfrac{(1+\beta)}{2\beta} \end{cases} \quad (8)$$

Hence, during an interaction between the pair, if we consider alternatively one observer placed on each of the two particles in a way such that all the energy and momentum of the DEMS is seen carried only at charge of the respective antiparticle, the equations (8) yields the description of the mutual interaction as an exchange of two de Broglie's photons with energy and momentum equal to the ones of one only of the interacting particles

$$\begin{cases} E_\gamma \approx \gamma m_0 c^2 \\ P_\gamma \approx \gamma\beta m_0 c \end{cases} \quad (9)$$

Equation (9) allows to consider the energy and momentum of the travelling particle as ones of the emitting wave in the direction of lab frame, so the two equations correspond to a de Broglie's wave description of a particle with frequency and wavelength:

$$\nu \approx \frac{\gamma m_0 c^2}{h} \ , \ \lambda \approx \frac{h}{\gamma\beta m_0 c} \quad (10)$$

but in this case the wave is an electromagnetic wave and not a probability wave.

This description can be used for each single particle $<i>$ that cross at high energy the polarizable frame S producing a kind of DEMS $\bullet SS_i$. In this case any external observer performs with its frame and the moving particle a dipolar source receiving a photon with an angle respect to the direction of the incoming particle that cannot be greater of a precise limit due light velocity. In order to define the emission angle in which the particle crossing the media S is seen to emit photons, considering the condition of existence of the eq. (7) we have:

$$\cos\theta < \frac{1}{\beta} \qquad (11)$$

always verified in vacuum where the source can emit the wave-particle in any direction, whereas for observers embedded in a polarizable material medium with a relative refraction index $n > 1$ eq. (11) gives for a travelling particle:

$$\theta \geq ar\cos\frac{1}{n\beta} \qquad (11a)$$

that describes for gamma emission the Cherenkov limit angle.

## 4) Conclusions

The Bridge Theory describes the physical world using a theoretical evolution of the electromagnetic interactions theory. Using the BT it is possible to have a consistent explanation of quantum phenomena and of relativistic effects without to use other theories. In fact, BT is able of to connect quantum mechanics and relativity in a way such to make them perfectly compatible.

To observe a charge particle we need of others charged particles to produce a mutual electromagnetic force. In fact only a charge feels a charge, in particular if we consider a charged particle that crosses a polarizable medium as neutral matter or a field of vacuum, the elementary charged particle interacts with at least a polarized anti-charge, producing a local DEMS. The source converts energy and momentum of the incoming particle in an electromagnetic wave. The energy and momentum exchanged between observer and the particle is the information concerning the dynamical relative state of the particle. When the DEMS is produced, the observer achieved in S by the wave measures the amounts of energy and the momentum of the wave like that of the particle. In this case the observation produces a phenomenology consistent with the de Broglie's wave. The wave with his energy and momentum has memory of the dynamical state of the particle that cross the medium.

In a real case when a particle crosses a polarizable medium, the source is not alone but a great number of DEMSs are produced and the energy and momentum of the particle are divided between each source. This produce in space a superposition of waves, i.e. the wave packet describing the particle. Each wave of the packet describes a peculiar state of the particle in relation with the DEMS produced along its path. Since the electromagnetic force acts to infinite, we can consider that along the path of the particle is produced an infinity of DEMSs with the effect to spread on the wave packet formed by the superposition of infinite waves all the energy and momentum of the incoming particle. The consequence of this peculiar interaction is to reduce to zero the energy and the momentum carried by each single wave, that is one of the infinite states describing the particle along its path. In this sense the electromagnetic wave associated to the DEMS production can be considered an empty wave that can be used to estimate the density of probability to realize the corresponding state.